\documentclass[prl,nobalancelastpage,twocolumn,nofootinbib,superscriptaddress,showpacs]{revtex4}
\usepackage{bm}
\usepackage{color}
\usepackage{amsfonts}
\usepackage{amssymb}
\usepackage{amsmath}
\usepackage{epsfig}
\usepackage{graphicx}
\usepackage{enumerate}
\usepackage{multirow}
\usepackage{verbatim}
\usepackage{color}
\usepackage{subfigure}

\begin{document}

\title{Infinite matrix product states, boundary conformal field theory, and
the open Haldane-Shastry model}

\author{Hong-Hao Tu}
\affiliation{Max-Planck Institut f\"ur Quantenoptik, Hans-Kopfermann-Str.~1, D-85748
Garching, Germany}
\author{Germ{\'a}n Sierra}
\affiliation{Instituto de F\'isica Te\'orica, UAM-CSIC, Madrid, Spain}
\affiliation{Department of Physics, Princeton University, Princeton, NJ 08544, USA}

\begin{abstract}
We show that infinite Matrix Product States (MPS) constructed from conformal
field theories can describe ground states of one-dimensional critical
systems with open boundary conditions. To illustrate this, we consider a
simple infinite MPS for a spin-1/2 chain and derive an inhomogeneous open
Haldane-Shastry model. For the spin-1/2 open Haldane-Shastry model, we
derive an exact expression for the two-point spin correlation function. We
also provide an SU($n$) generalization of the open Haldane-Shastry model and
determine its twisted Yangian generators responsible for the highly
degenerate multiplets in the energy spectrum.
\end{abstract}

\maketitle

\textit{Introduction.---} For a long time, it has been known that the main
curse of quantum many-body theory is the exponential growth of the Hilbert
space dimension with respect to the number of constituting particles. In the
last decades, the study of entanglement has significantly alleviated this
curse, at least to some extent, by recognizing the fact that only a tiny
corner of the Hilbert space, with small amount of entanglement, is pertinent
for the low-energy sector of Hamiltonians with local interactions. This deep
insight lies at the heart of tensor network states \cite{Verstraete08}, a
family of trial wave functions designed for efficiently representing the
physically relevant states in the tiny corner. The best known instance among
them is the Matrix Product States (MPS) in one spatial dimension, described
in terms of local matrices with finite dimensions. Their entanglement
entropies are bounded by the local matrix dimensions, which are nevertheless
sufficient for accurately approximating gapped ground states of
one-dimensional (1D) local Hamiltonians \cite{Verstraete06,Hastings07}. This
discovery not only provides a transparent theoretical picture for real-space
renormalization group methods \cite{Wilson75,White92}, but also leads to a
recent complete classification of all possible 1D gapped phases \cite%
{Pollmann10,Chen11,Norbert2011}.

For 1D critical systems, the low-energy physics is usually described by
conformal field theories (CFT). Their ground-state entanglement entropies
exhibit unbounded logarithmic growth \cite{Holzhey04,Vidal03,Calabrese04}
with respect to the subsystem size, indicating the deficiency of a usual MPS
description. To overcome this difficulty, infinite MPS, whose local matrices
are conformal fields living in an infinite-dimensional Hilbert space, have
been introduced in Ref.~\cite{Ignacio10}. The lattice sites for the infinite
MPS locate on a unit circle, embedded in a complex plane. This construction
shares conceptual similarity to Moore and Read's approach \cite{Moore91} of
writing 2D trial fractional quantum Hall states in terms of conformal
blocks. For a variety of examples \cite%
{Ignacio10,Anne11,Tu13,Tu14a,Tu14b,Bondesan14,Ivan14,Benedikt15}, the
infinite MPS (as well as their parent Hamiltonians) have been shown to
describe critical chains with \textit{periodic} boundary conditions (PBC)
and, furthermore, their critical behaviors are often related to the CFT
whose fields are used for constructing the wave functions \cite{exception}.
In this sense, the infinite MPS introduced in Ref.~\cite{Ignacio10} provide
a systematic way of finding lattice discretizations of CFT.

In this Rapid Communication, we show that the infinite MPS ansatz can
describe ground states of 1D critical systems with \textit{open} boundary
conditions (OBC), thus complementing the PBC case in Ref.~\cite{Ignacio10}.
Unlike \textit{bulk} CFT for periodic chains, open critical chains are
instead described by \textit{boundary} CFT. Taking a spin-1/2 chain as an
example, we show how the infinite MPS with an \textit{image} prescription
allows us to derive an inhomogeneous open Haldane-Shastry model, including
the original spin-1/2 open Haldane-Shastry models \cite{Simons94,Bernard95}
as special cases. Within the new formalism, an exact expression for the
two-point spin correlator of the spin-1/2 open Haldane-Shastry
model is obtained. This, together with numerical results for the
entanglement entropy, is in perfect agreement with the theoretical
predictions based on boundary CFT, which thus confirms that our infinite MPS
with the image prescription is suitable for describing open critical chains.
The open infinite MPS construction is readily applicable to any boundary CFT
for finding their lattice discretizations. As a further example, we derive
an SU($n$) generalization of the open Haldane-Shastry model. We characterize
its full spectrum and also determine the twisted Yangian generators
responsible for the highly degenerate multiplets in the energy spectrum.

\begin{figure}[tbp]
\centering
\includegraphics[width=0.95\linewidth]{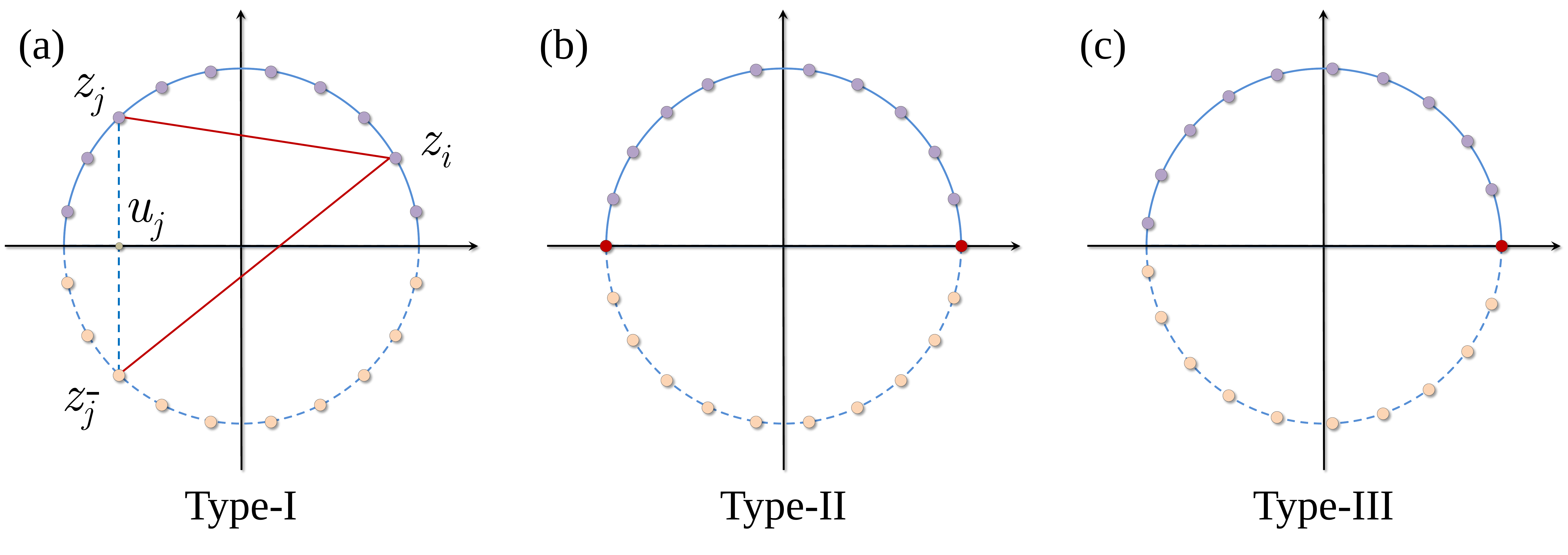}
\caption{(Color online) Schematic of an open chain in the upper complex
plane. The lattice sites and their mirror images locate on the upper and
lower unit semicircles, respectively. They are symmetric with respect to the
real axis. The two (brown) lines denote the chord distances $|z_{i}-z_{j}|$
and $|z_{i}-z_{\bar{j}}|$, respectively. (a)--(c) denote the three uniform
cases: (a) type-I: $\protect\theta _{j}=\frac{\protect\pi }{L}(j-\frac{1}{2}%
) $; (b) type-II: $\protect\theta _{j}=\frac{\protect\pi }{L+1}j$; (c)
type-III: $\protect\theta _{j}=\frac{2\protect\pi }{2L+1}j$.}
\label{fig:openchain}
\end{figure}

\textit{Infinite MPS and parent Hamiltonian.---} Let us consider a spin-1/2
chain located on the \textit{upper} unit circle in the complex plane, with $%
L $ lattice sites and complex lattice coordinates $z_{j}=e^{i\theta _{j}}$ ($%
j=1,\ldots ,L$ and $\theta _{j}\in \lbrack 0,\pi ]$ $\forall j$), see Fig.~%
\ref{fig:openchain}(a). We denote by $S_{j}^{a}$ ($a=1,2,3$)\ the spin-1/2
operators at site $j$. The local spin basis is defined by $|s_{j}\rangle $,
where $s_{j}=\pm 1$ (twice of the $S_{j}^{z}$ projection value). For each
site, we introduce its \textit{mirror} \textit{image} in the \textit{lower}
unit circle, e.g., site $j$ has an image $\bar{j}$, with complex coordinate $%
z_{\bar{j}}=z_{j}^{\ast }$. Following Ref.~\cite{Ignacio10}, the wave
function is written as a chiral correlator of CFT
fields:%
\begin{equation}
\Psi (s_{1},\ldots ,s_{L})=\langle A^{s_{1}}(u_{1})A^{s_{2}}(u_{2})\cdots
A^{s_{L}}(u_{L})\rangle ,  \label{eq:imps}
\end{equation}%
where $A^{s_{j}}(u_{j})=\chi _{j}:e^{is_{j}\phi (u_{j})/\sqrt{2}}:$ ($%
:\ldots :$ denotes normal ordering) and $u_{j}\equiv (z_{j}+z_{\bar{j}})/2$
(i.e., $u_{j}$ is the coordinate of the \textquotedblleft
barycenter\textquotedblright\ of $j$ and $\bar{j}$\ on the real axis). Here,
$\phi (u)$ is a chiral bosonic field from the $c=1$ free boson CFT, and $%
\chi _{j}=1,s_{j}$ for $j$ odd and even, respectively. Evaluating the chiral
correlator in (\ref{eq:imps}) yields a Jastrow wave function%
\begin{equation}
\Psi (s_{1},\ldots ,s_{L})=\delta _{s}e^{i\frac{\pi }{2}\sum_{i:\mathrm{even}%
}(s_{i}-1)}\prod_{j<l}(u_{j}-u_{l})^{s_{j}s_{l}/2},  \label{eq:jastrow}
\end{equation}%
where $\delta _{s}=1$ if $\sum_{j=1}^{L}s_{j}=0$ and zero otherwise (note
that $L$ must be even for ensuring a nonvanishing wave function). From the
explicit form (\ref{eq:jastrow}), it is transparent
that the sign factor (originated from $\chi _{j}$) is the
\textquotedblleft Marshall sign\textquotedblright , since the Jastrow
product in (\ref{eq:jastrow}) is positive.

As shown in Ref.~\cite{Ignacio10}, the infinite MPS (\ref{eq:imps}) with
coordinate choice $u_{j}=z_{j}=e^{i\frac{2\pi }{L}j}$, i.e., the case of
equidistantly distributed spins on the \textit{whole} unit circle, yields
the ground state of the SU(2) Haldane-Shastry model \cite%
{Haldane88,Shastry88}, which is a paradigmatic spin-1/2 chain with PBC.

Now we demonstrate that our infinite MPS (\ref{eq:imps}) with the image
prescription, $u_{j}\equiv (z_{j}+z_{\bar{j}})/2$, describes a spin-1/2
chain with OBC. Let us first derive a parent Hamiltonian for which (\ref%
{eq:jastrow}) is the exact ground state. Based on the CFT null field
techniques, it was shown \cite{Anne11}\ that the decoupling equations
satisfied by (\ref{eq:imps}) lead to a set of operators annihilating the
wave function (\ref{eq:jastrow}), $\mathcal{C}_{i}^{a}|\Psi \rangle =0$ $%
\forall i,a$, where $\mathcal{C}_{i}^{a}=\frac{2}{3}\sum_{j(\neq i)}\frac{1}{%
u_{i}-u_{j}}(S_{j}^{a}+i\varepsilon _{abc}S_{i}^{b}S_{j}^{c})$ and $\varepsilon _{abc}$ is the Levi-Civita symbol [we assume
summation over repeated indices and use the convention that $\sum_{j(\neq
i)} $ is the sum over $j$, whereas $\sum_{i\neq j}$ is the sum over both $i$
and $j$]. When adapting
to our present OBC setup, we consider the operators $\Lambda _{i}^{a}=\frac{2%
}{3}\sum_{j(\neq i)}(w_{ij}+w_{i\bar{j}})(S_{j}^{a}+i\varepsilon
_{abc}S_{i}^{b}S_{j}^{c})$, where $w_{ij}=(z_{i}+z_{j})/(z_{i}-z_{j})$ and
which also annihilate the wave function $|\Psi \rangle $, since $\Lambda
_{i}^{a}=(z_{i}-z_{i}^{\ast })\mathcal{C}_{i}^{a}$. The parent Hamiltonian
for (\ref{eq:jastrow}) is then defined as $H=\frac{1}{8}\sum_{i,a}(\Lambda
_{i}^{a})^{\dagger }\Lambda _{i}^{a}+\frac{L}{3}\mathbf{S}^{2}+E$, where $%
\mathbf{S}^{2}=\sum_{ij}\vec{S}_{i}\cdot \vec{S}_{j}$ is the total spin
operator and $E=\frac{1}{16}\sum_{i\neq j}(w_{ij}^{2}+w_{i\bar{j}}^{2})-%
\frac{1}{4}L^{2}$. After some algebra \cite{Supp}, we arrive at a long-range
Heisenberg model%
\begin{eqnarray}
H &=&\sum_{i\neq j}\left[ \frac{1}{|z_{i}-z_{j}|^{2}}+\frac{1}{|z_{i}-z_{%
\bar{j}}|^{2}}-\frac{w_{ij}(c_{i}-c_{j})}{12}\right.  \notag \\
&&\left. -\frac{w_{i\bar{j}}(c_{i}+c_{j})}{12}\right] (\vec{S}_{i}\cdot \vec{%
S}_{j})  \label{eq:openHS}
\end{eqnarray}%
with ground-state energy $E$, where $c_{j}=w_{\bar{j}j}+\sum_{l(\neq
j)}(w_{lj}+w_{\bar{l}j})$.

Three choices of the lattice coordinates deserve special attention (see Fig. %
\ref{fig:openchain}): (i) type-I: $\theta _{j}=\frac{\pi }{L}(j-\frac{1}{2})$%
; (ii) type-II: $\theta _{j}=\frac{\pi }{L+1}j$; (iii) type-III: $\theta
_{j}=\frac{2\pi }{2L+1}j$. For these three cases (termed as \textit{uniform}
cases afterwards), one obtains $w_{ij}(c_{i}-c_{j})+w_{i\bar{j}%
}(c_{i}+c_{j})=0$, $4$, and $2$, respectively. Accordingly, the parent
Hamiltonians, after removing the (unimportant) total spin operator $\mathbf{S%
}^{2}$ and constant terms in (\ref{eq:openHS}), have purely inverse-square
exchange interactions (between the spins and also their images), which
coincide with the open Haldane-Shastry models first introduced in Refs.~\cite%
{Simons94,Bernard95}. These uniform models are integrable and have highly
degenerate multiplets in their energy spectrum \cite{Simons94,Bernard95},
similar to their periodic counterpart \cite{Haldane92}, see Fig.~\ref%
{fig:spectrum} for the full spectrum of the open and periodic
Haldane-Shastry models with $L=6$. We postpone the discussion of this
degeneracy until presenting the SU($n$) generalization of these models,
where a unified treatment is possible. The Hamiltonian (\ref{eq:openHS})
with lattice coordinates other than the three uniform cases is an \textit{%
inhomogeneous} generalization of the open Haldane-Shastry models and does
not exhibit the huge degeneracy in the spectrum.

\begin{figure}[tbp]
\centering
\includegraphics[width=0.85\linewidth]{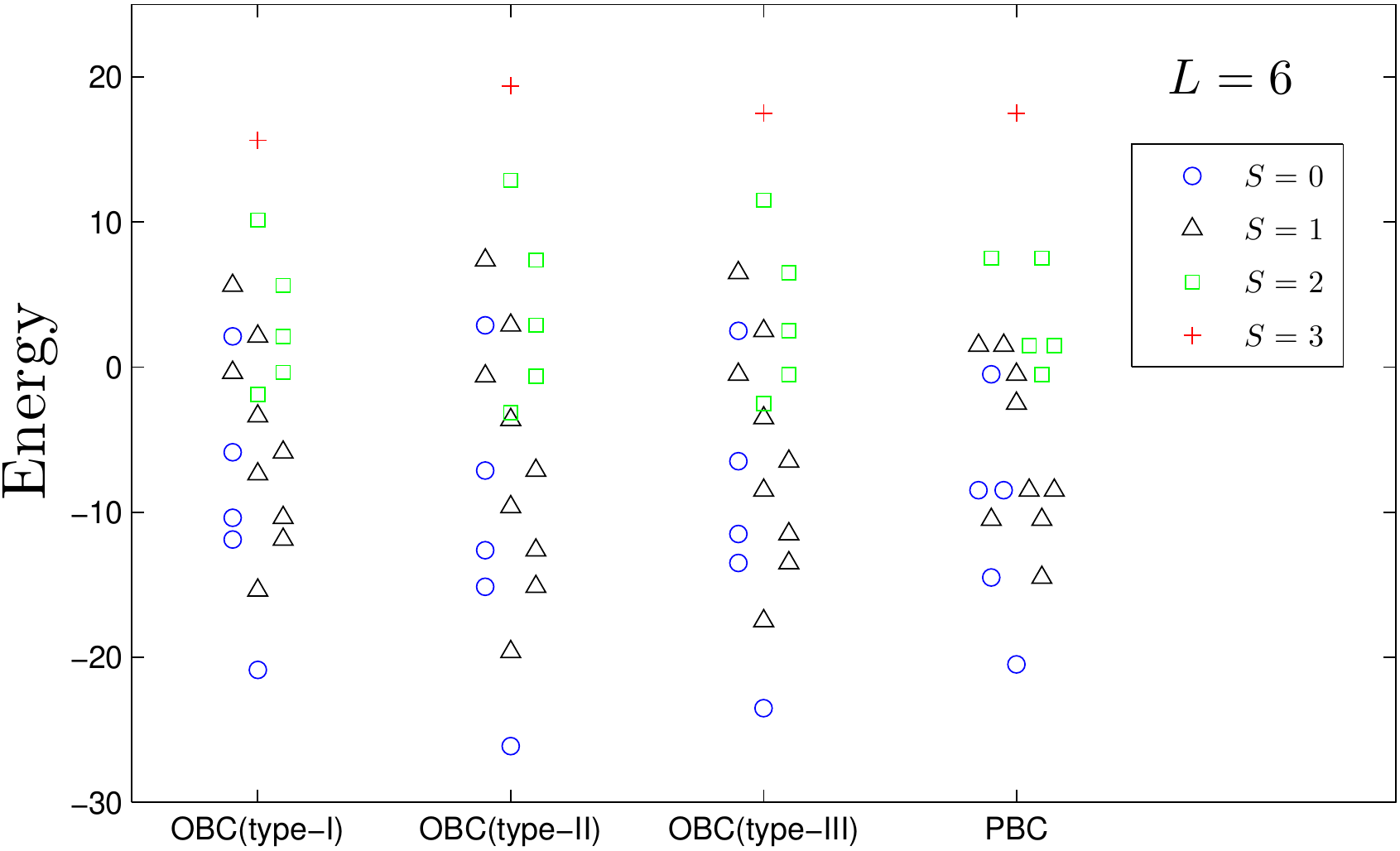}
\caption{(Color online) The energy spectrum of the three types of spin-1/2
open Haldane-Shastry models and the spin-1/2 periodic Haldane-Shastry model (%
$H=\sum_{i\neq j}\frac{\vec{S}_{i}\cdot \vec{S}_{j}}{\sin ^{2}[\protect\pi %
(i-j)/L]}$) with $L=6$. All four models have highly degenerate multiplets in
their energy spectrum. While the first excited states of the periodic model
are degenerate singlet and triplet (due to two free spin-1/2 spinons), the
open models do not have this degeneracy, indicating the importance of the
boundary effect.}
\label{fig:spectrum}
\end{figure}

\textit{Spin correlator.---} A nontrivial application of the infinite MPS
formulation is that, for the wave function (\ref{eq:jastrow}), the spin
correlation functions can be computed easily. Since $\mathcal{C}%
_{i}^{a}|\Psi \rangle =0$, one has $\langle \Psi |\sum_{a}S_{i}^{a}\mathcal{C%
}_{j}^{a}|\Psi \rangle =0$ and $\langle \Psi |\sum_{a}(\mathcal{C}%
_{j}^{a})^{\dagger }S_{i}^{a}|\Psi \rangle =0$ $\forall i,j$, which lead to
a set of linear equations relating two-point correlators $%
C_{ij}+\sum_{l(\neq i,j)}\frac{u_{i}-u_{j}}{u_{i}-u_{l}}C_{jl}=-\frac{3}{4}$
\cite{Anne11}, where $C_{ij}\equiv \langle \Psi |\vec{S}_{i}\cdot \vec{S}%
_{j}|\Psi \rangle /\langle \Psi |\Psi \rangle $. These equations are
sufficient for computing the two-point spin correlators for arbitrary
choices of $\theta _{j}$ (both inhomogeneous and uniform cases). The
generalization to arbitrary higher-order spin correlators is rather
straightforward.

Most remarkably, for the \textit{type-I} uniform case, these linear
equations allow us to find an analytical expression for the two-point spin
correlator \cite{Supp}%
\begin{eqnarray}
C_{ij} &=&\frac{3(-1)^{i-j}\sin \theta _{i}\sin \theta _{j}}{L(\cos \theta
_{i}-\cos \theta _{j})}\sum_{p=1}^{L/2}\sum_{q=0}^{p-1}g_{pq}[\cos
(2p-1)\theta _{i}  \notag \\
&&\times \cos 2q\theta _{j}-\cos 2q\theta _{i}\cos (2p-1)\theta _{j}]
\label{eq:correlator}
\end{eqnarray}%
with%
\begin{equation}
g_{pq}=\left\{
\begin{array}{c}
1 \\
\prod_{m=1}^{p-1}\frac{4m-1}{4m+1} \\
2\prod_{m=1}^{p-1}\frac{4m-1}{4m+1}\prod_{n=1}^{q}\frac{4n-3}{4n-1}%
\end{array}%
\right. \left.
\begin{array}{c}
p=1,q=0 \\
p>1,q=0 \\
p>1,q>0%
\end{array}%
\right. .  \label{eq:gpq}
\end{equation}%
\begin{figure}[tbp]
\centering
\includegraphics[width=0.95\linewidth]{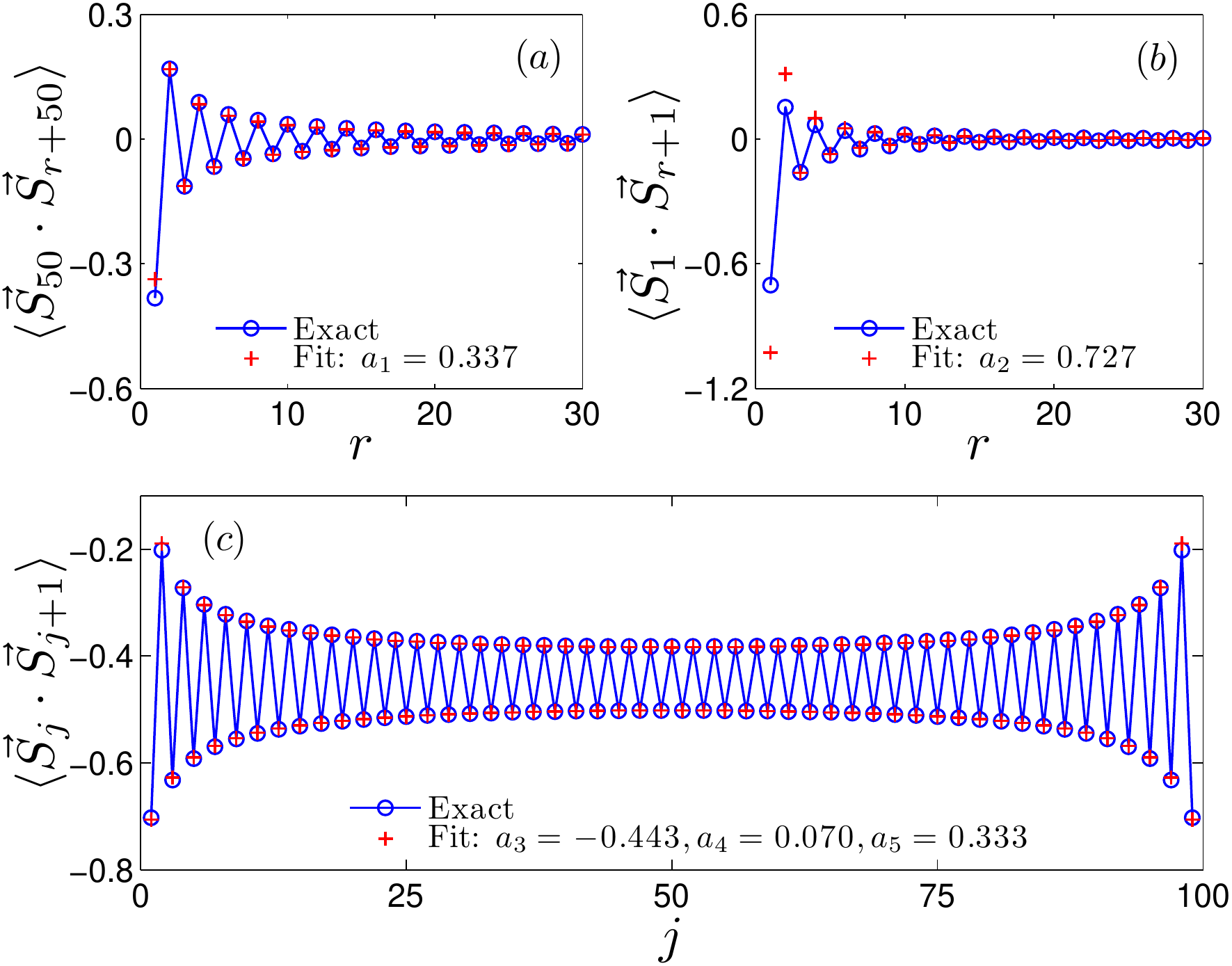}
\caption{(Color online) Two-point spin correlators of the wave function (%
\protect\ref{eq:jastrow}) in the type-I uniform case with $L=100$. The blue
circles are the exact results from (\protect\ref{eq:correlator}), and the
red crosses are fits with theoretical predictions based on the SU(2)$_{1}$
WZW model with free boundary condition (see text). (a) Two spins at lattice
sites $50$ and $50+r$ are far from the boundary. (b) One of the spins lives
at the boundary (the first spin). For (a) and (b), the first four points are
excluded when computing the fits, since the theoretical predictions are
valid for large $r$. (c) Two spins are nearest neighbors.}
\label{fig:correlator}
\end{figure}
In Fig.~\ref{fig:correlator} various correlators from (\ref{eq:correlator})
are compared with the theoretical predictions \cite{Eggert92} based on the
SU(2)$_{1}$ Wess-Zumino-Witten (WZW) model with free boundary condition.
When two spins at sites $j$ and $j+r$ are both far from the boundary, one
expects that the correlator $C_{j,j+r}$ recovers the result for PBC \cite%
{Gebhard87}, $C_{j,j+r}\simeq a_{1}(-1)^{r}/[\frac{2L}{\pi }\sin (\frac{\pi r%
}{2L})]\propto (-1)^{r}/r$ for large $r$, where $a_{1}$ is a constant.
However, if one of the two spins (say, the one at site $j$) is very close to
the boundary, the theory developed in Ref.~\cite{Eggert92} predicts $%
C_{j,j+r}\simeq a_{2}(-1)^{r}[\frac{L}{\pi }\sin (\frac{\pi r}{L})]^{1/2}/[%
\frac{2L}{\pi }\sin (\frac{\pi r}{2L})]^{2}\propto (-1)^{r}/r^{3/2}$ ($a_{2}$%
: nonuniversal constant) with\ a boundary critical exponent $\eta =3/2$ that
differs from $\eta =1$ in the bulk. For the correlator between nearest
neighbors, it was predicted \cite{Ng96,Laflorencie06} that $%
C_{j,j+1}=a_{3}+a_{4}/[\frac{L}{\pi }\sin (\frac{\pi j}{L}%
)]^{2}+a_{5}(-1)^{j}/[\frac{L}{\pi }\sin (\frac{\pi j}{L})]^{K}$, where $K$
is the Luttinger parameter, $K=1/2$, and $a_{3},a_{4},a_{5}$ are constants.
We treat the nonuniversal constants $a_{1},\ldots ,a_{5}$ as fitting
parameters and find excellent agreement between the exact result (\ref%
{eq:correlator}) and the SU(2)$_{1}$ WZW predictions (see Fig.~\ref%
{fig:correlator}).

\begin{figure}[tbp]
\centering
\includegraphics[width=0.95\linewidth]{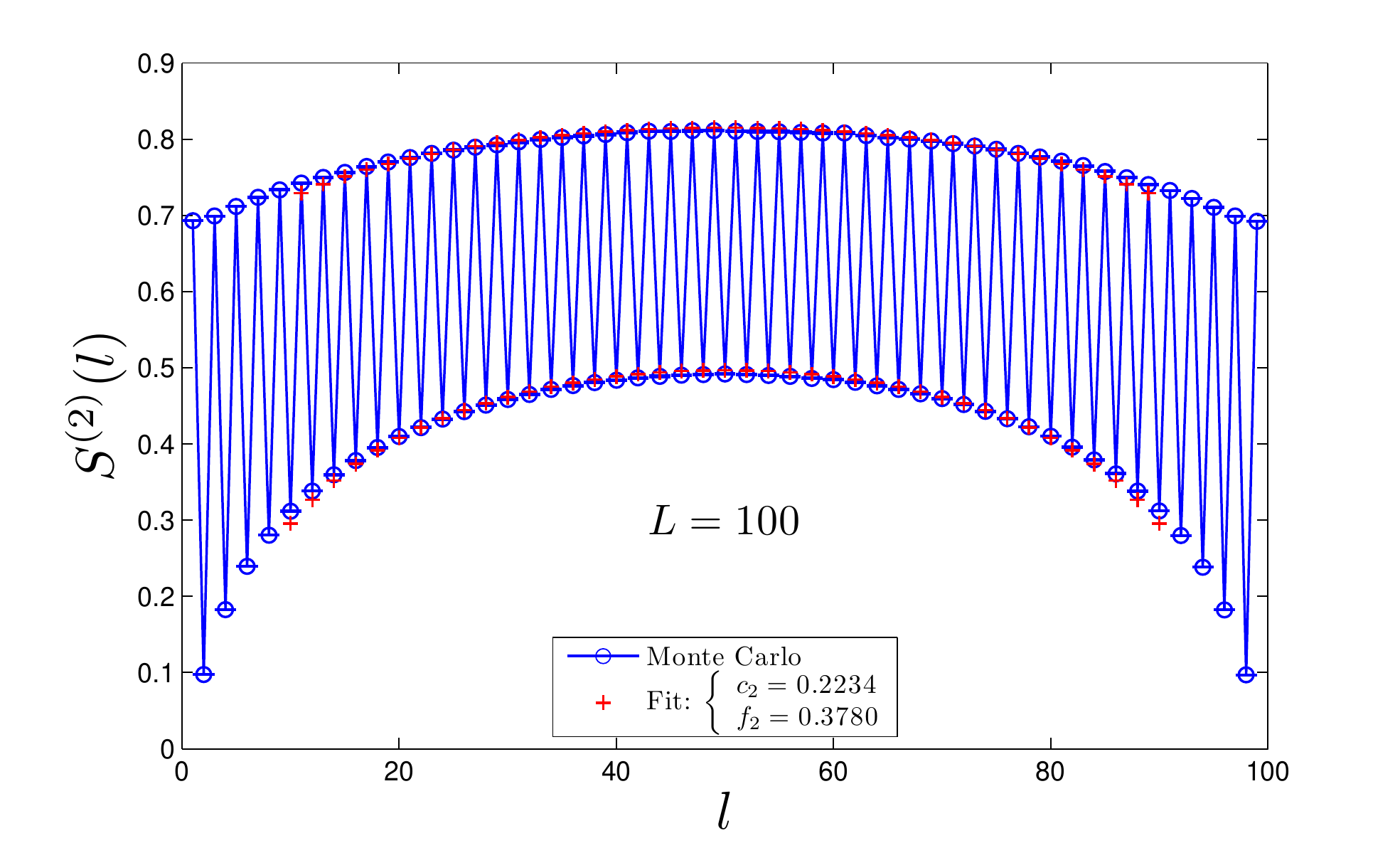}
\caption{(Color online) R\'{e}nyi entropy $S^{(2)}(l)$ of the wave function (%
\protect\ref{eq:jastrow}) in type-I uniform case with $L=100$ as a function
of the subsystem size $l$. The blue circles (with errorbars) are obtained
from Monte Carlo simulations and the red crosses are fits based on the
theoretical prediction (\protect\ref{eq:RenyiEntropy}) of the SU(2)$%
_{1}$ WZW model. The fit is computed with $l\in
\lbrack 10,90]$, as (\protect\ref{eq:RenyiEntropy}) is valid for large
subsystem sizes.}
\label{fig:entropy}
\end{figure}

\textit{Entanglement entropy.--- } To provide further support that the wave
function (\ref{eq:jastrow}) is relevant for open critical chains, we
numerically compute the R\'{e}nyi entropy $S^{(2)}(l)=-\ln \mathrm{Tr}\rho
_{l}^{2}$ via Monte Carlo method \cite{Ignacio10,Hastings10}, where $\rho
_{l}$ is the reduced density matrix of the first $l$ spins. In Fig.~\ref%
{fig:entropy}\ we plot $S^{(2)}(l)$ for the wave function (\ref{eq:jastrow})
in the type-I uniform case with $L=100$. For open spin-1/2 chains described
by the SU(2)$_{1}$ WZW model with free boundary condition, one expects the R%
\'{e}nyi entropy to be \cite{Laflorencie06}
\begin{equation}
S^{(2)}(l)=\frac{c}{8}\ln \left[ \frac{L}{\pi }\sin \left( \frac{\pi l}{L}%
\right) \right] +c_{2}+\frac{(-1)^{l}f_{2}}{[\frac{L}{\pi }\sin (\frac{\pi l%
}{L})]^{\frac{K}{2}}}  \label{eq:RenyiEntropy}
\end{equation}%
with central charge $c=1$, Luttinger parameter $K=1/2$, and $c_{2},f_{2}$
nonuniversal constants. Fixing $c=1$ and $K=1/2$ and treating $c_{2},f_{2}$
as fitting parameters, the numerical results are in good agreement with the
theoretical prediction (see Fig. \ref{fig:entropy}). For the type-II and
type-III uniform cases, we have verified via Monte Carlo simulations that
their R\'{e}nyi entropies also agree with (\ref{eq:RenyiEntropy}),
suggesting that they all belong to the SU(2)$_{1}$ WZW model with free
boundary condition.

\textit{SU(n) generalization.---} As a further application we generalize the
above SU(2) example to the SU($n$) case. For the SU($n$)$_{1}$ WZW model,
the infinite MPS have been proposed in Refs.~\cite{Tu14b,Bondesan14}. Here
we take in all sites SU($n$) spins transforming under \textit{fundamental}
representations, with local basis denoted by $|\alpha \rangle $ ($\alpha
=1,\ldots ,n$). Following Ref.~\cite{Tu14b}, the CFT fields for defining the
infinite MPS (\ref{eq:imps}) are given by $A^{\alpha }(u)=\kappa _{\alpha
}:e^{i\vec{m}_{\alpha }\cdot \vec{\phi}(u)/\sqrt{2}}:$, where $\vec{m}%
_{\alpha }$ is a $(n-1)$-component vector denoting the fundamental weight of
$|\alpha \rangle $ (e.g., $\vec{m}_{1,2}=(\pm 1,1/\sqrt{3})$ and $\vec{m}%
_{3}=(0,2/\sqrt{3})$ for SU(3), see \cite{Tu14b}), $\vec{\phi}(u)$ is a
vector of $n-1$ chiral bosonic fields, and $\kappa _{\alpha }$ is a Klein
factor, commuting with vertex operators and satisfying $\{\kappa _{\alpha
},\kappa _{\alpha ^{\prime }}\}=2\delta _{\alpha \alpha ^{\prime }}$.
Evaluating the CFT correlator (\ref{eq:imps}), the SU($n$) wave function
takes a simple Jastrow form, $\Psi _{\mathrm{SU(}n\mathrm{)}}(\alpha
_{1},\ldots ,\alpha _{L})=\mathrm{sgn}(x_{1}^{(1)},\ldots
,x_{L/n}^{(1)},\ldots ,x_{1}^{(n)},\ldots ,x_{L/n}^{(n)})\delta _{\sum_{i}%
\vec{m}_{\alpha _{i}}=0}\prod_{i<j}(u_{i}-u_{j})^{\delta _{\alpha _{i}\alpha
_{j}}}$ (sgn: signature of a permutation), where $x_{k}^{(\alpha )}$ ($%
k=1,\ldots ,L/n$), for a given configuration $|\alpha _{1},\ldots ,\alpha
_{L}\rangle $, is the position of the $k$th spin in the state $|\alpha
\rangle $.

Following a procedure similar to the SU(2) case \cite{Supp}, we obtain a
two-body parent Hamiltonian for $\Psi _{\mathrm{SU(}n)}$, $H=\sum_{i\neq j}%
\left[ \frac{1}{|z_{i}-z_{j}|^{2}}+\frac{1}{|z_{i}-z_{\bar{j}}|^{2}}-\frac{%
w_{ij}(c_{i}-c_{j})+w_{i\bar{j}}(c_{i}+c_{j})}{4(n+1)}\right] (\vec{t}%
_{i}\cdot \vec{t}_{j})$, where $t^{a}$ ($a=1,\ldots ,n^{2}-1$) are SU($n$)
generators in the fundamental representation, normalized as $\mathrm{tr}%
(t^{a}t^{b})=\frac{1}{2}\delta _{ab}$. The three uniform choices of $\theta
_{j}$, very much the same as the SU(2) cases, bring the parent Hamiltonian
into SU($n$) open Haldane-Shastry models
\begin{equation}
H_{\mathrm{SU(}n\mathrm{)}}=\sum_{i\neq j}\left[ \frac{1}{|z_{i}-z_{j}|^{2}}+%
\frac{1}{|z_{i}-z_{\bar{j}}|^{2}}\right] (\vec{t}_{i}\cdot \vec{t}_{j})
\label{eq:openSUnHS}
\end{equation}%
with purely inverse-square interactions.

Motivated by the SU(2) result \cite{Bernard95}, we have numerically observed
that the full spectrum of the SU($n$) open Haldane-Shastry model (\ref%
{eq:openSUnHS}) is described by the formula, $H_{\mathrm{SU(}n\mathrm{)}%
}|\{m_{i}\}\rangle =[E_{0}+\sum_{i=1}^{M}E(m_{i})]|\{m_{i}\}\rangle ,$ where
$E_{0}=-\frac{n-1}{8n}[\sum_{i\neq j}(w_{ij}^{2}+w_{i\bar{j}}^{2})-2L(L-1)]$
and $E(m_{i})=\frac{1}{2}(m_{i}^{2}-\frac{1}{4}N^{2})$ ($N=2L$, $2L+2$, and $%
2L+1$ for the three uniform cases, respectively), $M$ is an integer
satisfying $M\in \lbrack 0,\frac{n-1}{n}L]$, and $m_{i}$ are distinct
integer/half-integer rapidities ($1\leq m_{i}\leq L-1$, $2\leq m_{i}\leq L$,
and $\frac{3}{2}\leq m_{i}\leq L-\frac{1}{2}$ for each individual uniform
case), satisfying the generalized Pauli principle which is the same as that
for the SU($n$) Haldane-Shastry model with PBC \cite{Kawakami92,Ha92}: only
those sets $\{m_{1},\ldots ,m_{M}\}$ without $n$\ or more consecutive
integers/half-integers are allowed \cite{Haldane92}.

\textit{Twisted Yangian.---} Our numerical results also indicate that the
\textquotedblleft supermultiplet\textquotedblright\ structure in the
spectrum, which already shows up in the SU(2) case (see Fig.~\ref%
{fig:spectrum}), persists in the SU($n$) open Haldane-Shastry models (\ref%
{eq:openSUnHS}). To explain this degeneracy, we slightly generalize the
monodromy matrix found for the spin-1/2 open Haldane-Shastry models \cite%
{Bernard95} to the SU($n$) case. Through a third-order expansion of the
monodromy matrix \cite{Supp}, we obtain the nontrivial conserved charge
responsible for the SU($n$) open Haldane-Shastry models (\ref{eq:openSUnHS})
\begin{eqnarray}
Q^{a} &=&\sum_{k}t_{k}^{a}(w_{k\bar{k}}^{2}+\gamma _{1}w_{k0}^{2})-\gamma
_{2}\sum_{i\neq j\neq k}(w_{jk}+w_{j\bar{k}})  \notag \\
&&\times (w_{ij}-w_{i\bar{j}})t_{k}^{a}P_{jk}P_{ij},
\label{eq:twistedyangian}
\end{eqnarray}%
where $w_{k0}=(z_{k}+1)/(z_{k}-1)$, $P_{ij}$ swaps the spin states at site $%
i $ and $j$ (more explicitly, $P_{ij}=2\vec{t}_{i}\cdot \vec{t}_{j}+\frac{1}{%
n} $) and $\gamma _{1}$ and $\gamma _{2}$ are given by (i) type-I: $\gamma
_{1}=0$, $\gamma _{2}=\frac{1}{2}$; (ii) type-II: $\gamma _{1}=0$, $\gamma
_{2}=\frac{1}{10}$; (iii) type III: $\gamma _{1}=1$, $\gamma _{2}=\frac{1}{2}
$, respectively. The conserved charge $Q^{a}$ and the total spin $%
T^{a}\equiv \sum_{j}t_{j}^{a}$ both commute with (\ref{eq:openSUnHS}), but $%
Q^{a}$ does not commute with the SU($n$) Casimir operator $%
\sum_{a}T^{a}T^{a} $. This explains the appearance of degenerate eigenstates
with different SU($n$) representations. As the monodromy matrix relevant for
these models (with open boundaries) satisfies the reflection equation \cite%
{Sklyanin88}, the algebraic structure of the SU($n$) open Haldane-Shastry
models (\ref{eq:openSUnHS}) is the \textit{twisted Yangian} \cite%
{Olshanski92}. Thus, the conserved charges $Q^{a}$ and $T^{a}$ form the
lowest twisted Yangian generators.

\textit{Conclusions.---} In this Rapid Communication, we have shown that
infinite MPS with the image prescription are relevant for 1D critical chains
with OBC, by presenting a spin-1/2 example, as well as its SU($n$)
generalization. We have constructed inhomogeneous open Haldane-Shastry
models as their parent Hamiltonians, including the three open
Haldane-Shastry models as special uniform cases. For the type-I spin-1/2
open Haldane-Shastry model, an exact expression for the two-point spin
correlator has been derived and compared with theoretical
predictions, supporting that the low-energy effective theory is the SU(2)$%
_{1}$ WZW model with free boundary condition. We also characterize the full
spectrum of the SU($n$) open Haldane-Shastry models and determine the
twisted Yangian generators responsible for the highly degenerate multiplets
in the energy spectrum. The present infinite MPS with open boundaries is
readily applicable to any boundary CFT for finding their lattice
discretizations. As an outlook, we expect that the infinite MPS with OBC
could be very useful for proposing trial wave functions for single-impurity
Kondo problems, where boundary CFT are known \cite{Affleck90,Affleck91} to
play an important role.

\textit{Acknowledgment.---} We acknowledge J.~I.~Cirac and A.~E.~B.~Nielsen
for helpful discussions. This work has been supported by the EU project
SIQS, FIS2012-33642, QUITEMAD (CAM), the Severo Ochoa Program, and the
Fulbright grant PRX14/00352.

\onecolumngrid
\appendix
\setcounter{equation}{0} \newpage

\begin{center}
\textbf{Supplemental Material}
\end{center}

\section{Inhomogeneous open Haldane-Shastry models}

In this Section, we provide details on the derivation of the spin-1/2
inhomogeneous open Haldane-Shastry model and its SU($n$) generalization.

To construct the spin-1/2 inhomogeneous open Haldane-Shastry model, we use
the operators annihilating the spin-1/2 open infinite MPS%
\begin{equation}
\Lambda _{i}^{a}=\frac{2}{3}\sum_{j(\neq i)}(w_{ij}+w_{i\bar{j}%
})(S_{j}^{a}+i\varepsilon _{abc}S_{i}^{b}S_{j}^{c}),
\end{equation}%
to build a positive semidefinite operator%
\begin{eqnarray}
\sum_{a}(\Lambda _{i}^{a})^{\dagger }\Lambda _{i}^{a} &=&\frac{4}{9}%
\sum_{j,k(\neq i)}(w_{ij}^{\ast }+w_{i\bar{j}}^{\ast })(w_{ik}+w_{i\bar{k}%
})(S_{j}^{a}-i\varepsilon _{abc}S_{i}^{b}S_{j}^{c})(S_{k}^{a}+i\varepsilon
_{ade}S_{i}^{d}S_{k}^{e})  \notag \\
&=&-\frac{4}{9}\sum_{j,k(\neq i)}(w_{ij}+w_{i\bar{j}})(w_{ik}+w_{i\bar{k}})(%
\vec{S}_{j}\cdot \vec{S}_{k}-2i\varepsilon
_{abc}S_{i}^{a}S_{j}^{b}S_{k}^{c}+\varepsilon _{abc}\varepsilon
_{ade}S_{i}^{b}S_{i}^{d}S_{j}^{c}S_{k}^{e})  \notag \\
&=&-\frac{4}{9}\sum_{j,k(\neq i)}(w_{ij}+w_{i\bar{j}})(w_{ik}+w_{i\bar{k}})(%
\vec{S}_{j}\cdot \vec{S}_{k}-2i\varepsilon _{abc}S_{i}^{a}S_{j}^{b}S_{k}^{c}+%
\frac{1}{4}\varepsilon _{abc}\varepsilon _{abe}S_{j}^{c}S_{k}^{e}+\frac{i}{2}%
\varepsilon _{abc}\varepsilon _{ade}\varepsilon
_{bdf}S_{i}^{f}S_{j}^{c}S_{k}^{e})  \notag \\
&=&-\frac{2}{3}\sum_{j,k(\neq i)}(w_{ij}+w_{i\bar{j}})(w_{ik}+w_{i\bar{k}})(%
\vec{S}_{j}\cdot \vec{S}_{k}-i\varepsilon _{abc}S_{i}^{a}S_{j}^{b}S_{k}^{c})
\notag \\
&=&-\frac{2}{3}\sum_{j(\neq i)}(w_{ij}+w_{i\bar{j}})^{2}(\frac{3}{4}+\vec{S}%
_{i}\cdot \vec{S}_{j})-\frac{2}{3}\sum_{j\neq k(\neq i)}(w_{ij}+w_{i\bar{j}%
})(w_{ik}+w_{i\bar{k}})(\vec{S}_{j}\cdot \vec{S}_{k}),
\end{eqnarray}%
where we have used $w_{ij}^{\ast }=-w_{ij}$, $S^{b}S^{d}=\frac{1}{4}\delta
_{ab}+\frac{i}{2}\varepsilon _{abc}S^{c}$, $\varepsilon _{abc}\varepsilon
_{abd}=2\delta _{cd}$, and $\varepsilon _{abc}\varepsilon _{ade}\varepsilon
_{bdf}=\varepsilon _{cef}$. Then, we obtain%
\begin{equation}
\sum_{i,a}(\Lambda _{i}^{a})^{\dagger }\Lambda _{i}^{a}=-\frac{2}{3}%
\sum_{i\neq j}(w_{ij}+w_{i\bar{j}})^{2}(\frac{3}{4}+\vec{S}_{i}\cdot \vec{S}%
_{j})-\frac{2}{3}\sum_{j\neq k}\left( \sum_{i(\neq j,k)}(w_{ij}+w_{i\bar{j}%
})(w_{ik}+w_{i\bar{k}})\right) (\vec{S}_{j}\cdot \vec{S}_{k}).
\label{eq:PositiveOper}
\end{equation}

The following cyclic identity is the key for simplifying (\ref%
{eq:PositiveOper}):%
\begin{equation}
(w_{ij}+w_{i\bar{j}})(w_{ik}+w_{i\bar{k}})+(w_{ji}+w_{j\bar{\imath}%
})(w_{jk}+w_{j\bar{k}})+(w_{ki}+w_{k\bar{\imath}})(w_{kj}+w_{k\bar{j}})=4.
\label{eq:cyclic}
\end{equation}%
By using this identity, we obtain%
\begin{eqnarray}
\sum_{i(\neq j,k)}(w_{ij}+w_{i\bar{j}})(w_{ik}+w_{i\bar{k}}) &=&\sum_{i(\neq
j,k)}[4-(w_{ji}+w_{j\bar{\imath}})(w_{jk}+w_{j\bar{k}})-(w_{ki}+w_{k\bar{%
\imath}})(w_{kj}+w_{k\bar{j}})]  \notag \\
&=&4(L-2)-(w_{jk}+w_{j\bar{k}})\sum_{i(\neq j,k)}(w_{ji}+w_{j\bar{\imath}%
})-(w_{kj}+w_{k\bar{j}})\sum_{i(\neq j,k)}(w_{ki}+w_{k\bar{\imath}})  \notag
\\
&=&4(L-2)+2(w_{jk}^{2}+w_{j\bar{k}}^{2})-(w_{jk}+w_{j\bar{k}})\left[ w_{j%
\bar{j}}+\sum_{i(\neq j)}(w_{ji}+w_{j\bar{\imath}})\right]  \notag \\
&&-(w_{kj}+w_{k\bar{j}})\left[ w_{k\bar{k}}+\sum_{i(\neq k)}(w_{ki}+w_{k\bar{%
\imath}})\right] +w_{j\bar{j}}(w_{jk}+w_{j\bar{k}})+w_{k\bar{k}}(w_{kj}+w_{k%
\bar{j}})  \notag \\
&=&(4L-6)+2(w_{jk}^{2}+w_{j\bar{k}}^{2})+w_{jk}(c_{j}-c_{k})+w_{j\bar{k}%
}(c_{j}+c_{k}),  \label{eq:identity1}
\end{eqnarray}%
where we have defined $c_{j}\equiv w_{\bar{j}j}+\sum_{i(\neq j)}(w_{ij}+w_{%
\bar{\imath}j})$ and have used $w_{j\bar{j}}(w_{jk}+w_{j\bar{k}})+w_{k\bar{k}%
}(w_{kj}+w_{k\bar{j}})=2$ (the latter can be easily proved by using the
cyclic identity $w_{ij}w_{ik}+w_{ji}w_{jk}+w_{ki}w_{kj}=1$).

By substituting (\ref{eq:identity1}) into (\ref{eq:PositiveOper}), we arrive
at%
\begin{eqnarray}
\sum_{i,a}(\Lambda _{i}^{a})^{\dagger }\Lambda _{i}^{a} &=&-\frac{2}{3}%
\sum_{i\neq j}(w_{ij}+w_{i\bar{j}})^{2}(\frac{3}{4}+\vec{S}_{i}\cdot \vec{S}%
_{j})  \notag \\
&&-\frac{2}{3}\sum_{j\neq k}\left[ (4L-6)+2(w_{jk}^{2}+w_{j\bar{k}%
}^{2})+w_{jk}(c_{j}-c_{k})+w_{j\bar{k}}(c_{j}+c_{k})\right] (\vec{S}%
_{j}\cdot \vec{S}_{k})  \notag \\
&=&8\sum_{i\neq j}\left[ \frac{1}{|z_{i}-z_{j}|^{2}}+\frac{1}{|z_{i}-z_{\bar{%
j}}|^{2}}-\frac{w_{ij}(c_{i}-c_{j})+w_{i\bar{j}}(c_{i}+c_{j})}{12}\right] (%
\vec{S}_{i}\cdot \vec{S}_{j})  \notag \\
&&-\frac{8L}{3}\mathbf{S}^{2}-\frac{1}{2}\sum_{i\neq j}(w_{ij}+w_{i\bar{j}%
})^{2}+2L^{2},
\end{eqnarray}%
where we have used $w_{ij}^{2}=1-\frac{4}{|z_{i}-z_{j}|^{2}}$.

Then, the spin-1/2 inhomogeneous open Haldane-Shastry model is defined by%
\begin{eqnarray}
H &=&\frac{1}{8}\sum_{i,a}(\Lambda _{i}^{a})^{\dagger }\Lambda _{i}^{a}+%
\frac{L}{3}\mathbf{S}^{2}+E  \notag \\
&=&\sum_{i\neq j}\left[ \frac{1}{|z_{i}-z_{j}|^{2}}+\frac{1}{|z_{i}-z_{\bar{j%
}}|^{2}}-\frac{w_{ij}(c_{i}-c_{j})+w_{i\bar{j}}(c_{i}+c_{j})}{12}\right] (%
\vec{S}_{i}\cdot \vec{S}_{j}),  \label{eq:SU2H}
\end{eqnarray}%
whose ground-state energy $E$ is given by $E=\frac{1}{16}\sum_{i\neq
j}(w_{ij}+w_{i\bar{j}})^{2}-\frac{1}{4}L^{2}$.

The derivation of the SU($n$) inhomogeneous open Haldane-Shastry model
follows the similar steps for the spin-1/2 case. The operators annihilating
the SU($n$) infinite MPS are given by \cite{suppTu14b,suppBondesan14}
\begin{equation}
\Lambda _{i}^{a}=\frac{n+2}{2(n+1)}\sum_{j(\neq i)}(w_{ij}+w_{i\bar{j}%
})[t_{j}^{a}+(\frac{n}{n+2}d_{abc}+if_{abc})t_{i}^{b}t_{j}^{c}],
\end{equation}%
where $d_{abc}$ and $f_{abc}$ are the SU($n$) totally symmetry tensor and
the totally antisymmetric structure constant, respectively. Similar to the
spin-1/2 case, we consider the positive semidefinite operator%
\begin{eqnarray}
\sum_{a}(\Lambda _{i}^{a})^{\dagger }\Lambda _{i}^{a} &=&\frac{(n+2)^{2}}{%
4(n+1)^{2}}\sum_{j,k(\neq i)}(w_{ij}^{\ast }+w_{i\bar{j}}^{\ast
})(w_{ik}+w_{i\bar{k}})[t_{j}^{a}+(\frac{n}{n+2}%
d_{abc}+if_{abc})t_{i}^{b}t_{j}^{c}][t_{k}^{a}+(\frac{n}{n+2}%
d_{ade}+if_{ade})t_{i}^{d}t_{k}^{e}]  \notag \\
&=&\sum_{j,k(\neq i)}(w_{ij}^{\ast }+w_{i\bar{j}}^{\ast })(w_{ik}+w_{i\bar{k}%
})[\frac{n+2}{2(n+1)}(\vec{t}_{j}\cdot \vec{t}_{k})+\frac{n}{2(n+1)}%
d_{abc}t_{i}^{a}t_{j}^{b}t_{k}^{c}-\frac{n+2}{2(n+1)}%
if_{abc}t_{i}^{a}t_{j}^{b}t_{k}^{c}]  \notag \\
&=&-\sum_{j(\neq i)}(w_{ij}+w_{i\bar{j}})^{2}\left[ \frac{(n-1)(n+2)}{4n}+%
\frac{(n-1)(n+2)}{2(n+1)}(\vec{t}_{i}\cdot \vec{t}_{j})\right]  \notag \\
&&-\sum_{j\neq k(\neq i)}(w_{ij}+w_{i\bar{j}})(w_{ik}+w_{i\bar{k}})\left[
\frac{n+2}{2(n+1)}(\vec{t}_{j}\cdot \vec{t}_{k})+\frac{n}{2(n+1)}%
d_{abc}t_{i}^{a}t_{j}^{b}t_{k}^{c}\right] ,
\end{eqnarray}%
where we have extensively used the identities listed in the Appendix A in
Ref.~\cite{suppTu14b}. Notice that
\begin{eqnarray}
&&\sum_{i\neq j\neq k}(w_{ij}+w_{i\bar{j}})(w_{ik}+w_{i\bar{k}%
})d_{abc}t_{i}^{a}t_{j}^{b}t_{k}^{c}  \notag \\
&=&\frac{1}{3}\sum_{i\neq j\neq k}[(w_{ij}+w_{i\bar{j}})(w_{ik}+w_{i\bar{k}%
})+(w_{ji}+w_{j\bar{\imath}})(w_{jk}+w_{j\bar{k}})+(w_{ki}+w_{k\bar{\imath}%
})(w_{kj}+w_{k\bar{j}})]d_{abc}t_{i}^{a}t_{j}^{b}t_{k}^{c}  \notag \\
&=&\frac{4}{3}\sum_{i\neq j\neq k}d_{abc}t_{i}^{a}t_{j}^{b}t_{k}^{c}  \notag
\\
&=&\frac{4}{3}d_{abc}T^{a}T^{b}T^{c}-\frac{2(n^{2}-4)}{n}T^{a}T^{a}+\frac{%
2(n^{2}-1)(n^{2}-4)}{3n^{2}}L,
\end{eqnarray}%
where $T^{a}=\sum_{i}t_{i}^{a}$. Together with (\ref{eq:identity1}), we
obtain%
\begin{eqnarray}
\sum_{i,a}(\Lambda _{i}^{a})^{\dagger }\Lambda _{i}^{a} &=&-\sum_{i\neq
j}(w_{ij}+w_{i\bar{j}})^{2}\left[ \frac{(n-1)(n+2)}{4n}+\frac{(n-1)(n+2)}{%
2(n+1)}(\vec{t}_{i}\cdot \vec{t}_{j})\right]  \notag \\
&&-\sum_{i\neq j\neq k}(w_{ij}+w_{i\bar{j}})(w_{ik}+w_{i\bar{k}})\left[
\frac{n+2}{2(n+1)}(\vec{t}_{j}\cdot \vec{t}_{k})+\frac{n}{2(n+1)}%
d_{abc}t_{i}^{a}t_{j}^{b}t_{k}^{c}\right]  \notag \\
&=&2(n+2)\sum_{i\neq j}\left[ \frac{1}{|z_{i}-z_{j}|^{2}}+\frac{1}{|z_{i}-z_{%
\bar{j}}|^{2}}-\frac{w_{ij}(c_{i}-c_{j})+w_{i\bar{j}}(c_{i}+c_{j})}{4(n+1)}%
\right] (\vec{t}_{i}\cdot \vec{t}_{j})  \notag \\
&&-\frac{2n}{3(n+1)}d_{abc}T^{a}T^{b}T^{c}-\frac{2(n+2)L}{n+1}T^{a}T^{a}-%
\frac{(n-1)(n+2)}{4n}\sum_{i\neq j}(w_{ij}^{2}+w_{i\bar{j}}^{2}) \\
&&+\frac{(n-1)(n+2)}{6n}L(6L+n-2).  \notag
\end{eqnarray}

Then, the SU($n$) inhomogeneous open Haldane-Shastry model can be defined as%
\begin{eqnarray}
H &=&\frac{1}{2(n+2)}\sum_{i,a}(\Lambda _{i}^{a})^{\dagger }\Lambda _{i}^{a}+%
\frac{n}{3(n+1)(n+2)}d_{abc}T^{a}T^{b}T^{c}+\frac{L}{n+1}T^{a}T^{a}+E  \notag
\\
&=&\sum_{i\neq j}\left[ \frac{1}{|z_{i}-z_{j}|^{2}}+\frac{1}{|z_{i}-z_{\bar{j%
}}|^{2}}-\frac{w_{ij}(c_{i}-c_{j})+w_{i\bar{j}}(c_{i}+c_{j})}{4(n+1)}\right]
(\vec{t}_{i}\cdot \vec{t}_{j}),  \label{eq:SUnH}
\end{eqnarray}%
whose ground-state energy $E$ is given by $E=\frac{n-1}{8n}\sum_{i\neq
j}(w_{ij}^{2}+w_{i\bar{j}}^{2})-\frac{n-1}{12n}L(6L+n-2)$.

\section{Two-point spin correlation function for the type-I spin-1/2 open
Haldane-Shastry model}

In this Section, we derive the exact expression of the two-point spin
correlation function for the \textit{type-I} spin-1/2 open Haldane-Shastry
model.

As we mentioned in the main text, the two-point spin correlation function $%
C_{ij}=\langle \Psi |\vec{S}_{i}\cdot \vec{S}_{j}|\Psi \rangle /\langle \Psi
|\Psi \rangle $ satisfies the following linear equations:%
\begin{equation}
\frac{1}{u_{i}-u_{j}}C_{ij}+\sum_{l(\neq i,j)}\frac{1}{u_{i}-u_{l}}C_{jl}=-%
\frac{3}{4}\frac{1}{u_{i}-u_{j}},\text{ \ }\forall i,j
\end{equation}%
where $u_{j}=\cos \theta _{j}$. Since $|\Psi \rangle $ is a spin singlet, $%
\sum_{j=1}^{L}\vec{S}_{j}|\Psi \rangle =0$, the correlator also satisfies%
\begin{equation}
\sum_{j(\neq i)}C_{ij}=-\frac{3}{4}.
\end{equation}

For instance, if one wants to determine the correlators involving the first
spin, one could write down the $L-1$ linear equations (relating $C_{1j}$, $%
j=2,\ldots ,L$) in a matrix form:%
\begin{equation}
\begin{pmatrix}
-\frac{1}{u_{1}-u_{2}} & \frac{1}{u_{2}-u_{3}} & \frac{1}{u_{2}-u_{4}} &
\cdots & \frac{1}{u_{2}-u_{L}} \\
\frac{1}{u_{3}-u_{2}} & -\frac{1}{u_{1}-u_{3}} & \frac{1}{u_{3}-u_{4}} &
\cdots & \frac{1}{u_{3}-u_{L}} \\
\frac{1}{u_{4}-u_{2}} & \frac{1}{u_{4}-u_{3}} & -\frac{1}{u_{1}-u_{4}} &
\cdots & \frac{1}{u_{4}-u_{L}} \\
\vdots & \vdots & \vdots & \ddots & \vdots \\
\frac{1}{u_{L}-u_{2}} & \frac{1}{u_{L}-u_{3}} & \frac{1}{u_{L}-u_{4}} &
\cdots & -\frac{1}{u_{1}-u_{L}}%
\end{pmatrix}%
\begin{pmatrix}
C_{12} \\
C_{13} \\
C_{14} \\
\vdots \\
C_{1L}%
\end{pmatrix}%
=-\frac{3}{4}%
\begin{pmatrix}
\frac{1}{u_{2}-u_{1}} \\
\frac{1}{u_{3}-u_{1}} \\
\frac{1}{u_{4}-u_{1}} \\
\vdots \\
\frac{1}{u_{L}-u_{1}}%
\end{pmatrix}%
.  \label{eq:MatrixEquation}
\end{equation}%
The correlators involving other spins can be solved in a similar fashion.
For the moment, we carry out the derivations based on (\ref%
{eq:MatrixEquation}) for ease of notation and, in the end, extend the final
result to the most general case.

For the type-I case with $\theta _{j}=\frac{\pi }{L}(j-\frac{1}{2})$, the
following sum identity is very useful:%
\begin{equation}
\sum_{j(\neq i)}\frac{1}{u_{j}-u_{i}}\cos m\theta _{j}=\frac{2(L-m)\sin
\theta _{i}\sin m\theta _{i}-\cos \theta _{i}\cos m\theta _{i}}{2\sin
^{2}\theta _{i}},  \label{eq:sum}
\end{equation}%
where $m$ is an integer and $m\in \lbrack 0,2L]$.

For the $l$-th row in (\ref{eq:MatrixEquation}), we multiply $\cos m\theta
_{l+1}$ and then sum over all the linear equations. By using (\ref{eq:sum}),
we obtain%
\begin{equation}
\sum_{j=2}^{L}\left[ \frac{\cos m\theta _{1}+\cos m\theta _{j}}{\cos \theta
_{1}-\cos \theta _{j}}-\frac{2(L-m)\sin \theta _{j}\sin m\theta _{j}-\cos
\theta _{j}\cos m\theta _{j}}{2\sin ^{2}\theta _{j}}\right] C_{1j}=\frac{3}{8%
}\frac{2(L-m)\sin \theta _{1}\sin m\theta _{1}-\cos \theta _{1}\cos m\theta
_{1}}{\sin ^{2}\theta _{1}},  \label{eq:Linear1}
\end{equation}%
where $m\in \lbrack 0,2L]$. For $m=0$, this yields%
\begin{equation}
\sum_{j=2}^{L}\left( \frac{2}{\cos \theta _{1}-\cos \theta _{j}}+\frac{\cos
\theta _{j}}{2\sin ^{2}\theta _{j}}\right) C_{1j}=-\frac{3}{8}\frac{\cos
\theta _{1}}{\sin ^{2}\theta _{1}}.  \label{eq:Linear2}
\end{equation}%
When multiplying (\ref{eq:Linear2}) by $\cos m\theta _{1}$ and then
subtracting with (\ref{eq:Linear1}), we obtain%
\begin{equation}
\sum_{j=2}^{L}\left[ \frac{\cos m\theta _{j}-\cos m\theta _{1}}{\cos \theta
_{1}-\cos \theta _{j}}-\frac{2(L-m)\sin \theta _{j}\sin m\theta _{j}+\cos
\theta _{j}(\cos m\theta _{1}-\cos m\theta _{j})}{2\sin ^{2}\theta _{j}}%
\right] C_{1j}=\frac{3}{4}\frac{(L-m)\sin m\theta _{1}}{\sin \theta _{1}}.
\label{eq:Linear3}
\end{equation}%
Manipulating three consecutive linear equations [taking $m-1$, $m$, and $m+1$
in (\ref{eq:Linear3})], we arrive at%
\begin{equation}
\sum_{j(\neq 1)}\left[ (2L-2m+1)\frac{\cos (m+1)\theta _{j}}{\sin ^{2}\theta
_{j}}-(2L-2m-1)\frac{\cos (m-1)\theta _{j}}{\sin ^{2}\theta _{j}}\right]
(\cos \theta _{1}-\cos \theta _{j})C_{1j}=3\cos m\theta _{1},
\label{eq:Linear4}
\end{equation}%
which we have verified to hold for $m\in \lbrack 0,2L]$.

In general, the two-point spin correlator satisfies the following equation:%
\begin{equation}
\sum_{j(\neq i)}\left[ (2L-2m+1)\frac{\cos (m+1)\theta _{j}}{\sin ^{2}\theta
_{j}}-(2L-2m-1)\frac{\cos (m-1)\theta _{j}}{\sin ^{2}\theta _{j}}\right]
(\cos \theta _{i}-\cos \theta _{j})C_{ij}=3\cos m\theta _{i},
\label{eq:Linear5}
\end{equation}%
where $m\in \lbrack 0,2L]$.

In practice, finding the analytical form of $C_{ij}$ directly from (\ref%
{eq:Linear5}) does not seem to be a simple task. Here we adopt an approach
used in Ref.~\cite{suppKuramotoBook} to determine the analytical form of $%
C_{ij}$ for a few finite-size chains, from which a well-educated guess helps
to solve (\ref{eq:Linear5}).

In the hardcore boson basis, the type-I open Haldane-Shastry ground state is
written as%
\begin{equation}
|\Psi \rangle =\sum_{x_{1}<\ldots <x_{L/2}}\Psi (x_{1},\ldots
,x_{L/2})S_{x_{1}}^{+}\cdots S_{x_{L/2}}^{+}|0\rangle ,
\label{eq:hardcoreboson}
\end{equation}%
where%
\begin{equation}
\Psi (x_{1},\ldots
,x_{L/2})=(-1)^{\sum_{i=1}^{L/2}x_{i}}\prod_{l=1}^{L/2}\sin \theta
_{x_{l}}\prod_{1\leq i<j\leq L/2}(\cos \theta _{x_{i}}-\cos \theta
_{x_{j}})^{2}.
\end{equation}%
Here $x_{1},\ldots ,x_{L/2}$ denote the positions of the hardcore bosons (up
spins).

The norm of (\ref{eq:hardcoreboson}) is given by%
\begin{eqnarray}
\langle \Psi |\Psi \rangle &=&\sum_{x_{1}<\ldots <x_{L/2}}|\Psi
(x_{1},\ldots ,x_{L/2})|^{2}  \notag \\
&=&\frac{1}{(L/2)!}\sum_{x_{1},\ldots ,x_{L/2}}\prod_{l=1}^{L/2}\sin
^{2}\theta _{x_{l}}\prod_{1\leq i<j\leq L/2}(\cos \theta _{x_{i}}-\cos
\theta _{x_{j}})^{4}  \notag \\
&=&\frac{1}{(L/2)!}\sum_{x_{1},\ldots ,x_{L/2}}\prod_{l=1}^{L/2}\sin
^{2}\theta _{x_{l}}\det
\begin{pmatrix}
1 & \cos \theta _{x_{1}} & \cos ^{2}\theta _{x_{1}} & \cos ^{3}\theta
_{x_{1}} & \cdots & \cos ^{L-1}\theta _{x_{1}} \\
0 & 1 & 2\cos \theta _{x_{1}} & 3\cos ^{2}\theta _{x_{1}} & \cdots &
(L-1)\cos ^{L-2}\theta _{x_{1}} \\
1 & \cos \theta _{x_{2}} & \cos ^{2}\theta _{x_{2}} & \cos ^{3}\theta
_{x_{2}} & \cdots & \cos ^{L-1}\theta _{x_{2}} \\
0 & 1 & 2\cos \theta _{x_{2}} & 3\cos ^{2}\theta _{x_{2}} & \cdots &
(L-1)\cos ^{L-2}\theta _{x_{2}} \\
\vdots & \vdots & \vdots & \vdots & \ddots & \vdots \\
1 & \cos \theta _{x_{L/2}} & \cos ^{2}\theta _{x_{L/2}} & \cos ^{3}\theta
_{x_{L/2}} & \cdots & \cos ^{L-1}\theta _{x_{L/2}} \\
0 & 1 & 2\cos \theta _{x_{L/2}} & 3\cos ^{2}\theta _{x_{L/2}} & \cdots &
(L-1)\cos ^{L-2}\theta _{x_{L/2}}%
\end{pmatrix}%
,  \label{eq:norm}
\end{eqnarray}%
where in the last step we have used the \textit{Confluent Alternant}
identity \cite{suppKuramotoBook}
\begin{equation}
\prod_{1\leq i<j\leq M}(y_{i}-y_{j})^{4}=\det
\begin{pmatrix}
1 & y_{1} & y_{1}^{2} & y_{1}^{3} & \cdots & y_{1}^{M-1} \\
0 & 1 & 2y_{1} & 3y_{1}^{2} & \cdots & (M-1)y_{1}^{M-2} \\
1 & y_{2} & y_{2}^{2} & y_{2}^{3} & \cdots & y_{2}^{M-1} \\
0 & 1 & 2y_{2} & 3y_{2}^{2} & \cdots & (M-1)y_{2}^{M-2} \\
\vdots & \vdots & \vdots & \vdots & \ddots & \vdots \\
1 & y_{M} & y_{M}^{2} & y_{M}^{3} & \cdots & y_{M}^{M-1} \\
0 & 1 & 2y_{M} & 3y_{M}^{2} & \cdots & (M-1)y_{M}^{M-2}%
\end{pmatrix}%
.  \label{eq:ConfluentAlternant}
\end{equation}

Similarly, the unnormalized transverse spin correlator (for $i\neq j$) can
be expressed as%
\begin{eqnarray}
\langle \Psi |S_{i}^{+}S_{j}^{-}|\Psi \rangle &=&\frac{1}{(L/2-1)!}%
\sum_{x_{1},\ldots ,x_{L/2-1}}\Psi ^{\ast }(i,x_{1},\ldots ,x_{L/2-1})\Psi
(j,x_{1},\ldots ,x_{L/2-1})  \notag \\
&=&\frac{-(-1)^{i-j}}{(L/2-1)!}\frac{\sin \theta _{i}\sin \theta _{j}}{\cos
\theta _{i}-\cos \theta _{j}}\sum_{x_{1},\ldots
,x_{L/2-1}}\prod_{l=1}^{L/2-1}\sin ^{2}\theta _{x_{l}}  \notag \\
&&\times \det
\begin{pmatrix}
1 & \cos \theta _{i} & \cos ^{2}\theta _{i} & \cos ^{3}\theta _{i} & \cdots
& \cos ^{L-1}\theta _{i} \\
1 & \cos \theta _{j} & \cos ^{2}\theta _{j} & \cos ^{3}\theta _{j} & \cdots
& \cos ^{L-1}\theta _{j} \\
1 & \cos \theta _{x_{1}} & \cos ^{2}\theta _{x_{1}} & \cos ^{3}\theta
_{x_{1}} & \cdots & \cos ^{L-1}\theta _{x_{1}} \\
0 & 1 & 2\cos \theta _{x_{1}} & 3\cos ^{2}\theta _{x_{1}} & \cdots &
(L-1)\cos ^{L-2}\theta _{x_{1}} \\
\vdots & \vdots & \vdots & \vdots & \ddots & \vdots \\
1 & \cos \theta _{x_{L/2-1}} & \cos ^{2}\theta _{x_{L/2-1}} & \cos
^{3}\theta _{x_{L/2-1}} & \cdots & \cos ^{L-1}\theta _{x_{L/2-1}} \\
0 & 1 & 2\cos \theta _{x_{L/2-1}} & 3\cos ^{2}\theta _{x_{L/2-1}} & \cdots &
(L-1)\cos ^{L-2}\theta _{x_{L/2-1}}%
\end{pmatrix}%
.  \label{eq:TransverseCorrelator}
\end{eqnarray}

For small $L$, (\ref{eq:norm}) and (\ref{eq:TransverseCorrelator}) can be
computed by expanding the determinants (with Laplace's formula). After the
expansion, the discrete sums over the coordinates can be carried out by
using the following identities:%
\begin{equation}
\sum_{x=1}^{L}\sin ^{2}\theta _{x}\cos ^{2r}\theta _{x}=\frac{1}{r+1}\frac{1%
}{2^{2r+1}}\binom{2r}{r}L,
\end{equation}%
and%
\begin{equation}
\sum_{x=1}^{L}\sin ^{2}\theta _{x}\cos ^{2r+1}\theta _{x}=0,
\end{equation}%
which are valid for the type-I case and $r=0,\ldots ,\frac{L}{2}-1$.

Following this procedure, we obtain for $L=4$%
\begin{eqnarray}
\frac{\langle \Psi |S_{i}^{+}S_{j}^{-}|\Psi \rangle }{\langle \Psi |\Psi
\rangle } &=&\frac{(-1)^{i-j}}{L}\frac{\sin \theta _{i}\sin \theta _{j}}{%
\cos \theta _{i}-\cos \theta _{j}}[2(\cos \theta _{i}-\cos \theta _{j})+%
\frac{6}{5}(\cos 3\theta _{i}-\cos 3\theta _{j})  \notag \\
&&-\frac{4}{5}(\cos 2\theta _{i}\cos 3\theta _{j}-\cos 3\theta _{i}\cos
2\theta _{j})].  \label{eq:L4}
\end{eqnarray}%
For $L=6$, we obtain%
\begin{eqnarray}
\frac{\langle \Psi |S_{i}^{+}S_{j}^{-}|\Psi \rangle }{\langle \Psi |\Psi
\rangle } &=&\frac{(-1)^{i-j}}{L}\frac{\sin \theta _{i}\sin \theta _{j}}{%
\cos \theta _{i}-\cos \theta _{j}}[2(\cos \theta _{i}-\cos \theta _{j})+%
\frac{6}{5}(\cos 3\theta _{i}-\cos 3\theta _{j})+\frac{14}{15}(\cos 5\theta
_{i}-\cos 5\theta _{j})  \notag \\
&&-\frac{4}{5}(\cos 2\theta _{i}\cos 3\theta _{j}-\cos 3\theta _{i}\cos
2\theta _{j})-\frac{28}{45}(\cos 2\theta _{i}\cos 5\theta _{j}-\cos 5\theta
_{i}\cos 2\theta _{j})  \notag \\
&&-\frac{4}{9}(\cos 4\theta _{i}\cos 5\theta _{j}-\cos 5\theta _{i}\cos
4\theta _{j})].  \label{eq:L6}
\end{eqnarray}%
For $L=8$, we obtain%
\begin{eqnarray}
\frac{\langle \Psi |S_{i}^{+}S_{j}^{-}|\Psi \rangle }{\langle \Psi |\Psi
\rangle } &=&\frac{(-1)^{i-j}}{L}\frac{\sin \theta _{i}\sin \theta _{j}}{%
\cos \theta _{i}-\cos \theta _{j}}[2(\cos \theta _{i}-\cos \theta _{j})+%
\frac{6}{5}(\cos 3\theta _{i}-\cos 3\theta _{j})+\frac{14}{15}(\cos 5\theta
_{i}-\cos 5\theta _{j})  \notag \\
&&+\frac{154}{195}(\cos 7\theta _{i}-\cos 7\theta _{j})-\frac{4}{5}(\cos
2\theta _{i}\cos 3\theta _{j}-\cos 3\theta _{i}\cos 2\theta _{j})-\frac{28}{%
45}(\cos 2\theta _{i}\cos 5\theta _{j}-\cos 5\theta _{i}\cos 2\theta _{j})
\notag \\
&&-\frac{308}{585}(\cos 2\theta _{i}\cos 7\theta _{j}-\cos 7\theta _{i}\cos
2\theta _{j})-\frac{4}{9}(\cos 4\theta _{i}\cos 5\theta _{j}-\cos 5\theta
_{i}\cos 4\theta _{j})  \notag \\
&&-\frac{44}{117}(\cos 4\theta _{i}\cos 7\theta _{j}-\cos 7\theta _{i}\cos
4\theta _{j})-\frac{4}{13}(\cos 6\theta _{i}\cos 7\theta _{j}-\cos 7\theta
_{i}\cos 6\theta _{j})].  \label{eq:L8}
\end{eqnarray}

Since $\langle \Psi |S_{i}^{+}S_{j}^{-}|\Psi \rangle =\langle \Psi
|S_{i}^{-}S_{j}^{+}|\Psi \rangle $ and $|\Psi \rangle $ is a spin singlet,
we have%
\begin{equation}
C_{ij}=\frac{\langle \Psi |\vec{S}_{i}\cdot \vec{S}_{j}|\Psi \rangle }{%
\langle \Psi |\Psi \rangle }=\frac{3}{2}\frac{\langle \Psi
|S_{i}^{+}S_{j}^{-}|\Psi \rangle }{\langle \Psi |\Psi \rangle }.
\end{equation}

For larger $L$, a direct computation of (\ref{eq:norm}) and (\ref%
{eq:TransverseCorrelator}) becomes quickly involved. However, from the
finite-size results (\ref{eq:L4})--(\ref{eq:L8}), there is an indication
that, for general $L$, the analytical form of the two-point spin correlator $%
C_{ij}$ is given by%
\begin{equation}
C_{ij}=\frac{3(-1)^{i-j}\sin \theta _{i}\sin \theta _{j}}{L(\cos \theta
_{i}-\cos \theta _{j})}\sum_{p=1}^{L/2}\sum_{q=0}^{p-1}g_{pq}[\cos
(2p-1)\theta _{i}\cos 2q\theta _{j}-\cos 2q\theta _{i}\cos (2p-1)\theta
_{j}],  \label{eq:corr}
\end{equation}%
where $g_{pq}$ has no $L$ dependence and its initial values are readily
available from (\ref{eq:L8}).

By substituting (\ref{eq:corr}) into (\ref{eq:Linear5}), the well-educated
guess (\ref{eq:corr}) indeed solves the linear equation and the general
expression for $g_{pq}$ is found to be
\begin{equation}
g_{pq}=\left\{
\begin{array}{c}
1 \\
\prod_{m=1}^{p-1}\frac{4m-1}{4m+1} \\
2\prod_{m=1}^{p-1}\frac{4m-1}{4m+1}\prod_{n=1}^{q}\frac{4n-3}{4n-1}%
\end{array}%
\right. \left.
\begin{array}{c}
p=1,q=0 \\
p>1,q=0 \\
p>1,q>0%
\end{array}%
\right. .
\end{equation}

\section{Twisted Yangian generators for the SU($n$) open Haldane-Shastry
model}

In this Section, we provide details on the derivation of the twisted Yangian
generators for the SU($n$) open Haldane-Shastry model.

For the SU(2) open Haldane-Shastry model, such formalism has already been
developed in Ref.~\cite{suppBernard95}. Althought its SU($n$) generalization
is rather straightforward, we present the derivation below for the purpose
of being self-contained.

Following Ref.~\cite{suppBernard95}, we introduce an \textit{unprojected}
Hamiltonian%
\begin{equation}
\hat{H}=-\sum_{i\neq j}\left[ \frac{z_{i}z_{j}}{(z_{i}-z_{j})^{2}}(K_{ij}-1)+%
\frac{z_{i}z_{j}^{-1}}{(z_{i}-z_{j}^{-1})^{2}}(\bar{K}_{ij}-1)\right]
-\sum_{i=1}^{L}\left[ b_{1}\frac{z_{i}}{(z_{i}-1)^{2}}+2b_{2}\frac{1}{%
(z_{i}-z_{i}^{-1})^{2}}\right] (K_{i}-1),  \label{eq:unprojectedH}
\end{equation}%
where the coordinates $z_{i}$ are viewed as \textit{dynamical} variables,
the coordinate permutation operators $K_{ij}$, $\bar{K}_{ij}$, and $K_{i}$,
when acting on the coordinates, yield $K_{ij}z_{i}=z_{j}K_{ij}$, $\bar{K}%
_{ij}z_{i}=z_{j}^{-1}\bar{K}_{ij}$, and $K_{i}z_{i}=z_{i}^{-1}K_{i}$, and
the constants $b_{1}$ and $b_{2}$ will be specified below.

We also define a \textit{projection }operation $\pi $ which replaces the
operators $K_{ij}$ and $\bar{K}_{ij}$ by the SU($n$) spin permutation
operator $P_{ij}=2\vec{t}_{i}\cdot \vec{t}_{j}+\frac{1}{2}$, and $K_{i}$ by
the identity operator once they have been moved to the right of an
expression. In the simplest case with only one of these operators, we have%
\begin{eqnarray}
\pi (K_{ij}) &=&\pi (\bar{K}_{ij})=P_{ij}, \\
\pi (K_{i}) &=&1.
\end{eqnarray}%
If there are multiply coordinate permutation operators $K_{ij}$ and $\bar{K}%
_{ij}$ present, the rule of the projection operation is to insert a designed
product of SU($n$) spin permutation operators (which itself should be an
identity, e.g., $P_{ik}P_{ij}P_{ij}P_{ik}=1$) into the expression and then
replace each combined product $P_{ij}K_{ij}$ (appearing to the right of an
expression) by an identity, e.g.,%
\begin{equation}
\pi (K_{ij}K_{ik})=\pi (P_{ik}P_{ij}P_{ij}K_{ij}P_{ik}K_{ik})=P_{ik}P_{ij}.
\end{equation}

After the projection operation, the coordinates are \textit{not} dynamical
any more. Then, the \textit{projected} Hamiltonian is a pure SU($n$) spin
model%
\begin{eqnarray}
H &=&\pi (\hat{H})  \notag \\
&=&-\sum_{i\neq j}\left[ \frac{z_{i}z_{j}}{(z_{i}-z_{j})^{2}}(P_{ij}-1)+%
\frac{z_{i}z_{j}^{-1}}{(z_{i}-z_{j}^{-1})^{2}}(P_{ij}-1)\right]  \notag \\
&=&\sum_{i\neq j}\left[ \frac{1}{|z_{i}-z_{j}|^{2}}+\frac{1}{|z_{i}-z_{\bar{j%
}}|^{2}}\right] (P_{ij}-1).  \label{eq:projectedH}
\end{eqnarray}

In Ref.~\cite{suppBernard95}, it has been shown that the projected
Hamiltonian is integrable, if the lattice coordinates correspond to the
three uniform cases (see Fig. 1 in the main text) and the constants $b_{1}$
and $b_{2}$ in (\ref{eq:unprojectedH})\ are given by (i) type-I: $b_{1}=0$
and $b_{2}=1$; (ii) type-II: $b_{1}=0$ and $b_{2}=3$; (iii) type-III: $%
b_{1}=b_{2}=1$. Notice that the three projected Hamiltonians (\ref%
{eq:projectedH}), after subtracting a constant, just correspond to the open
SU($n$) Haldane-Shastry model [Eq.~(7) in the main text].

The integrability becomes manifest by introducing the Dunkl operators%
\begin{equation}
d_{i}=\sum_{j(>i)}\frac{z_{i}}{z_{i}-z_{j}}K_{ij}-\sum_{j(<i)}\frac{z_{j}}{%
z_{i}-z_{j}}K_{ij}+\sum_{j(\neq i)}\frac{z_{i}}{z_{i}-z_{j}^{-1}}\bar{K}%
_{ij}+\left( b_{1}\frac{z_{i}}{z_{i}-1}+b_{2}\frac{z_{i}}{z_{i}-z_{i}^{-1}}%
\right) K_{i},  \label{eq:dunkl}
\end{equation}%
which are mutually commuting, $[d_{i},d_{j}]=0$ $\forall i,j$, and all
commute with the unprojected Hamiltonian, $[d_{i},\hat{H}]=0$ $\forall i$.
After introducing an extra $n$-dimensional auxiliary Hilbert space (denoted
by \textquotedblleft $0$\textquotedblright ), the SU($n$) monodromy matrix $%
T(u)$ can be defined as%
\begin{equation}
T(u)=\pi \left[ \prod_{i=1}^{L}\left( 1+\frac{P_{i0}}{u-d_{i}}\right) \left(
1+\frac{b_{1}+b_{2}}{2}\frac{1}{u}\right) \prod_{i=L}^{1}\left( 1+\frac{%
P_{i0}}{u+d_{i}}\right) \right] ,  \label{eq:monodromy}
\end{equation}%
which is a $n\times n$ operator-valued matrix function of the spectral
parameter $u$. Actually, it is a generating function of conserved charges, $%
[T(u),H]=0$ $\forall u$. By using the Taylor expansion $\frac{1}{u-d_{i}}=%
\frac{1}{u}+\frac{d_{i}}{u^{2}}+\frac{d_{i}^{2}}{u^{3}}+\mathcal{O}(1/u^{4})$
and implementing the projection, one obtains formally the following
expression:
\begin{equation}
T(u)=1+\frac{1}{u}\left( t_{0}^{0}\otimes
J_{1}^{0}+\sum_{a=1}^{n^{2}-1}t_{0}^{a}\otimes J_{1}^{a}\right) +\frac{1}{%
u^{2}}\left( t_{0}^{0}\otimes J_{2}^{0}+\sum_{a=1}^{n^{2}-1}t_{0}^{a}\otimes
J_{2}^{a}\right) +\cdots ,
\end{equation}%
where $J_{\mu }^{0}$ and $J_{\mu }^{a}$ ($a=1,\ldots ,n^{2}-1$ and $\mu
=1,\ldots ,\infty $) are conserved charges for the SU($n$) open
Haldane-Shastry model, $[J_{\mu }^{0},H]=[J_{\mu }^{a},H]=0$. For the
monodromy matrix (\ref{eq:monodromy}), the conserved charges in the first-
and secord-order expansions in $1/u$ are trivial (such as $T^{a}$, $%
d_{abc}T^{b}T^{c}$, $T^{a}T^{a}$, etc). In the third-order expansion, we
obtain, after a tedious but straightforward calculation, the following
nontrivial conserved charge:%
\begin{equation}
Q^{a}=\sum_{k}t_{k}^{a}(w_{k\bar{k}}^{2}+\gamma _{1}w_{k0}^{2})-\gamma
_{2}\sum_{i\neq j\neq k}(w_{jk}+w_{j\bar{k}})(w_{ij}-w_{i\bar{j}%
})t_{k}^{a}P_{jk}P_{ij},
\end{equation}%
where $\gamma _{1}$ and $\gamma _{2}$, for the three uniform cases, are
given by (i) type-I: $\gamma _{1}=0$, $\gamma _{2}=\frac{1}{2}$; (ii)
type-II: $\gamma _{1}=0$, $\gamma _{2}=\frac{1}{10}$; (iii) type-III: $%
\gamma _{1}=1$, $\gamma _{2}=\frac{1}{2}$, respectively.


\begin{thebibliography}{99}
\bibitem{Verstraete08} F.~Verstraete, V.~Murg, and J.~I.~Cirac, \textit{%
Matrix product states, projected entangled pair states, and variational
renormalization group methods for quantum spin systems}, Adv. Phys. \textbf{%
57}, 143 (2008).

\bibitem{Verstraete06} F.~Verstraete and J.~I.~Cirac, \textit{Matrix product
states represent ground states faithfully}, Phys. Rev. B \textbf{73}, 094423
(2006).

\bibitem{Hastings07} M.~B.~Hastings, \textit{An area law for one-dimensional
quantum systems}, J. Stat. Mech. (2007) P08024.

\bibitem{Wilson75} K.~G.~Wilson, \textit{The renormalization group: Critical
phenomena and the Kondo problem}, Rev. Mod. Phys. \textbf{47}, 773 (1975).

\bibitem{White92} S.~R.~White, \textit{Density matrix formulation for
quantum renormalization groups}, Phys. Rev. Lett. \textbf{69}, 2863 (1992).

\bibitem{Pollmann10} F.~Pollmann, A.~M.~Turner, E.~Berg, and M.~Oshikawa,
\textit{Entanglement spectrum of a topological phase in one dimension},
Phys. Rev. B \textbf{81}, 064439 (2010).

\bibitem{Chen11} X.~Chen, Z.-C. Gu, and X.-G.~Wen, \textit{Classification of
gapped symmetric phases in one-dimensional spin systems}, Phys. Rev. B
\textbf{83}, 035107 (2011).

\bibitem{Norbert2011} N.~Schuch, D.~Perez-Garcia, and J.~I.~Cirac, \textit{%
Classifying quantum phases using matrix product states and projected
entangled pair state}s, Phys. Rev. B \textbf{84}, 165139 (2011).

\bibitem{Holzhey04} C.~Holzhey, F.~Larsen, and F.~Wilczek, \textit{Geometric
and renormalized entropy in conformal field theory}, Nucl. Phys. B \textbf{%
424}, 443 (1994).

\bibitem{Vidal03} G.~Vidal, J.~I.~Latorre, E.~Rico, and A.~Kitaev, \textit{%
Entanglement in quantum critical phenomena}, Phys. Rev. Lett. \textbf{90},
227902 (2003).

\bibitem{Calabrese04} P.~Calabrese and J.~Cardy, \textit{Entanglement
entropy and quantum field theory}, J. Stat. Mech. (2004) P06002.

\bibitem{Ignacio10} J.~I.~Cirac and G.~Sierra, \textit{Infinite matrix
product states, conformal field theory, and the Haldane-Shastry model},
Phys. Rev. B \textbf{81}, 104431 (2010).

\bibitem{Moore91} G.~Moore and N.~Read, \textit{Nonabelions in the
fractional quantum Hall effect}, Nucl. Phys. B \textbf{360}, 362 (1991).

\bibitem{Anne11} A.~E.~B.~Nielsen, J.~I.~Cirac, and G.~Sierra, \textit{%
Quantum spin Hamiltonians for the SU(2)}$_{k}$\textit{\ WZW model}, J. Stat.
Mech. (2011) P11014.

\bibitem{Tu13} H.-H.~Tu, \textit{Projected BCS states and spin Hamiltonians
for the SO(}$n$\textit{)}$_{1}$\textit{\ Wess-Zumino-Witten model}, Phys.
Rev. B \textbf{87}, 041103 (2013).

\bibitem{Tu14a} H.-H.~Tu, A.~E.~B.~Nielsen, J.~I.~Cirac, and G.~Sierra,
\textit{Lattice Laughlin states of bosons and fermions at filling fractions }%
$1/q$, New J. Phys. \textbf{16}, 033025 (2014).

\bibitem{Tu14b} H.-H.~Tu, A.~E.~B.~Nielsen, and G.~Sierra, \textit{Quantum
spin models for the SU(}$n$\textit{)}$_{1}$\textit{\ Wess-Zumino-Witten model%
}, Nucl. Phys. B \textbf{886}, 328 (2014).

\bibitem{Bondesan14} R.~Bondesan and T.~Quella, \textit{Infinite matrix
product states for long-range SU(}$N$\textit{) spin models}, Nucl. Phys. B
\textbf{886}, 483 (2014).

\bibitem{Ivan14} I.~Glasser, J.~I.~Cirac, G.~Sierra, and A.~E.~B.~Nielsen,
\textit{Construction of spin models displaying quantum criticality from
quantum field theory}, Nucl. Phys. B \textbf{886}, 63 (2014).

\bibitem{Benedikt15} B.~Herwerth, G.~Sierra, H.-H.~Tu, and A.~E.~B.~Nielsen,
\textit{Excited states in spin chains from conformal blocks},
arXiv:1501.07557.

\bibitem{exception} Note however that exceptional cases exist, for which the
connection of the critical behaviors of the infinite MPS and the CFT for
constructing them is unclear, see, e.g., the SU($n$) states with alternating
fundamental and conjugate representations in Refs.~\cite{Tu14b,Bondesan14}.

\bibitem{Simons94} B.~D.~Simons and B.~L.~Altshuler, \textit{Exact ground
state of an open }$S=1/2$\textit{\ long-range Heisenberg antiferromagnetic
spin chain}, Phys. Rev. B \textbf{50}, 1102 (1994).

\bibitem{Bernard95} D.~Bernard, V.~Pasquier, and D.~Serban, \textit{Exact
solution of long-range interacting spin chains with boundaries}, Europhys.
Lett. \textbf{30}, 301 (1995).

\bibitem{Haldane88} F.~D.~M.~Haldane, \textit{Exact Jastrow-Gutzwiller
resonating-valence-bond ground state of the spin-1/2 antiferromagnetic
Heisenberg chain with }$1/r^{2}$\textit{\ exchange}, Phys. Rev. Lett.
\textbf{60}, 635 (1988).

\bibitem{Shastry88} B.~S.~Shastry, \textit{Exact solution of an }$S=1/2$%
\textit{\ Heisenberg antiferromagnetic chain with long-ranged interactions},
Phys. Rev. Lett. \textbf{60}, 639 (1988).

\bibitem{Supp} See Supplemental Material for the derivations of the SU(2)
inhomogeneous open Haldane-Shastry model and its SU($n$) generalization, the
two-point spin correlation function for the type-I SU(2) open
Haldane-Shastry model, and the twisted Yangian generators for the SU($n$)
open Haldane-Shastry model, which includes Ref.~\cite{KuramotoBook}.

\bibitem{KuramotoBook} Y.~Kuramoto and Y.~Kato, \textit{Dynamics of
one-dimensional quantum systems: inverse-square interaction models}
(Cambridge University Press, New York, 2009).

\bibitem{Haldane92} F.~D.~M.~Haldane, Z.~N.~C.~Ha, J.~C.~Talstra,
D.~Bernard, and V. Pasquier, \textit{Yangian symmetry of integrable quantum
chains with long-range interactions and a new description of states in
conformal field theory}, Phys. Rev. Lett. \textbf{69}, 2021 (1992).

\bibitem{Eggert92} S.~Eggert and I.~Affleck, \textit{Magnetic impurities in
half-integer-spin Heisenberg antiferromagnetic chains}, Phys. Rev. B \textbf{%
46}, 10866 (1992).

\bibitem{Gebhard87} F. Gebhard and D. Vollhardt, \textit{Correlation
functions for Hubbard-type models: The exact results for the Gutzwiller wave
function in one dimension}, Phys. Rev. Lett. \textbf{59}, 1472 (1987).

\bibitem{Ng96} T.-K.~Ng, S.-J.~Qin, and Z.-B.~Su, \textit{Density-matrix
renormalization-group study of }$S=1/2$\textit{\ Heisenberg spin chains:
Friedel oscillations and marginal system-size effects}, Phys. Rev. B \textbf{%
54}, 9854 (1996).

\bibitem{Laflorencie06} N.~Laflorencie, E.~S.~S\o rensen, M.-S.~Chang, and
I.~Affleck, \textit{Boundary Effects in the Critical Scaling of Entanglement
Entropy in 1D Systems}, Phys. Rev. Lett. \textbf{96}, 100603 (2006).

\bibitem{Hastings10} M.~B.~Hastings, I.~Gonz\'{a}lez, A.~B.~Kallin, and
R.~G. Melko, \textit{Measuring Renyi Entanglement Entropy in Quantum Monte
Carlo Simulations}, Phys. Rev. Lett. \textbf{104}, 157201 (2010).

\bibitem{Kawakami92} N.~Kawakami, \textit{Asymptotic Bethe-ansatz solution
of multicomponent quantum systems with }$1/r^{2}$\textit{\ long-range
interaction}, Phys. Rev. B \textbf{46}, 1005 (1992).

\bibitem{Ha92} Z.~N.~C.~Ha and F.~D.~M.~Haldane, \textit{Models with
inverse-square exchange}, Phys. Rev. B \textbf{46}, 9359 (1992).


\bibitem{Sklyanin88} E.~K.~Sklyanin, \textit{Boundary conditions for
integrable quantum systems}, J. Phys. A \textbf{21}, 2375 (1998).

\bibitem{Olshanski92} G.~I.~Olshanski, \textit{Twisted Yangians and
infinite-dimensional classical Lie algebras}, Quantum Groups (edited by P.
P. Kulish), Lecture Notes in Math. \textbf{1510}, (Springer, Berlin, 1992).

\bibitem{Affleck90} I.~Affleck, \textit{A current algebra approach to the
Kondo effect}, Nucl. Phys. B \textbf{336}, 517 (1990).

\bibitem{Affleck91} I.~Affleck and A.~W.~W.~Ludwig, \textit{The Kondo
effect, conformal field theory and fusion rules}, Nucl. Phys. B \textbf{352}%
, 849 (1991); \textit{Critical theory of overscreened Kondo fixed points},
Nucl. Phys. B \textbf{360}, 641 (1991).
\end{thebibliography}

\begin{thebibliography}{9}
\bibitem{suppTu14b} H.-H.~Tu, A.~E.~B.~Nielsen, and G.~Sierra, \textit{%
Quantum spin models for the SU(}$n$\textit{)}$_{1}$\textit{\
Wess-Zumino-Witten model}, Nucl. Phys. B \textbf{886}, 328 (2014).

\bibitem{suppBondesan14} R.~Bondesan and T.~Quella, \textit{Infinite matrix
product states for long-range SU(}$N$\textit{) spin models}, Nucl. Phys. B
\textbf{886}, 483 (2014).

\bibitem{suppKuramotoBook} Y.~Kuramoto and Y.~Kato, \textit{Dynamics of
one-dimensional quantum systems: inverse-square interaction models}
(Cambridge University Press, New York, 2009).

\bibitem{suppBernard95} D.~Bernard, V.~Pasquier, and D.~Serban, \textit{%
Exact solution of long-range interacting spin chains with boundaries},
Europhys. Lett. \textbf{30}, 301 (1995).
\end{thebibliography}
\end{document}